\documentclass[conference]{IEEEtran}
\usepackage{cite}
\usepackage{amsmath,amssymb,amsfonts}
\usepackage{algorithmic}
\usepackage{graphicx}
\usepackage{textcomp}
\usepackage{xcolor}
\usepackage[utf8]{inputenc}
\usepackage{comment}
\usepackage{makecell}
\usepackage{cleveref}
\usepackage{tablefootnote}
\usepackage{adjustbox}

\usepackage[labelformat=simple, labelsep=colon]{subcaption}
\DeclareCaptionSubType*{figure}

\usepackage[english, shorthands=off]{babel}
\usepackage[left=1.57cm,right=1.57cm,top=1.3cm,bottom=2.54cm]{geometry}

\def\BibTeX{{\rm B\kern-.05em{\sc i\kern-.025em b}\kern-.08em
    T\kern-.1667em\lower.7ex\hbox{E}\kern-.125emX}}
\begin{document}

\newcommand{\bigsizee}{\fontsize{16pt}{24pt}\selectfont}

\title{Improving PKI, BGP, and DNS Using Blockchain: A Systematic Review
\\\bigsizee{Parça Zinciri ile PKI, BGP ve DNS İyileştirmeleri: Sistematik Bir İnceleme}
}

\author{\IEEEauthorblockN{Faizan Safdar Ali, Alptekin Küpçü}
\IEEEauthorblockA{\textit{Computer Science and Engineering, Koç University, İstanbul, TURKEY}\\
\{fali18,akupcu\}@ku.edu.tr}
}

\maketitle

\renewcommand{\abstractname}{Abstract}
\begin{abstract}  
The Internet has many backbone components on top of which the whole world is connected. It is important to make these components, like Border Gateway Protocol (BGP), Domain Name System (DNS), and Public Key Infrastructure (PKI), secure and work without any interruption. All of the aforementioned components have vulnerabilities, mainly because of their dependence on the centralized parties, that should be resolved. 

Blockchain is revolutionizing the concept of today's Internet, primarily because of its degree of decentralization and security properties. In this paper, we discuss how blockchain provides nearly complete solutions to the open challenges for these network backbone components.	
\end{abstract}
\renewcommand\IEEEkeywordsname{Keywords}
\begin{IEEEkeywords}
Blockchain, Internet, BGP, DNS, PKI.
\end{IEEEkeywords}

\renewcommand{\abstractname}{Öz}
\begin{abstract}
Dünya çapında bağlantı sağlayan İnternet çeşitli omurga bileşenlere sahiptir. Sınır Geçidi Protokolü (BGP), Alan Adı Sistemi (DNS) ve Açık Anahtar Altyapısı (PKI) gibi bileşenlerin güvenli hale getirilerek ve kesintisiz çalışmalarının sağlanması. Bu bileşenlerin özellikle merkezi otoritelere olan bağımlılıkları nedeniyle çözülmesi gereken zayıf noktaları vardır.

Parça zinciri dağıtık çalışma ve güvenlik özellikleri nedeniyle günümüzün İnternet kavramında devrim yaratan bir yapıdır. Bu makalede, parça zincirinin belirtilen omurga bileşenlerdeki sıkıntılara nasıl neredeyse bütüncül çözümler sunduğunu tartışıyoruz.
\end{abstract}
\renewcommand\IEEEkeywordsname{Anahtar Sözcükler}
\begin{IEEEkeywords}
	Parça Zinciri, Blokzincir, İnternet, BGP, DNS, PKI.
\end{IEEEkeywords}

\section{Introduction}
The first design of the Internet was presented as a centralized single entity. With time, the Internet was divided into sub-systems (e.g., DNS, BGP, PKI). Still, these components were built on a centralized architecture. This introduces security, privacy, and performance issues such as a single point of failure, trust issues, high latencies, and storage \cite{b30}. This results in these centralized services getting hacked frequently. Efforts were put in to make the services distributed \cite{b29}. These improved the Internet by solving the above-mentioned challenges but introduced new types of issues like scheduling, resource allocation, coordination, device management, scalability, security, trust, and multiple weak points of contact for attackers \cite{b29}. The overall performance was decreased because of the replication of work, backups, and the communication of distributed parts of the overall system \cite{b29}. Recently, blockchain-based solutions were introduced, with the goal being improving security while keeping the speed, cost and correctness comparable to the legacy components presented in Table \ref{LPP}. We can see that the current protocols take less time and cost (as most of the vendors have already implemented them) and achieve the desired (correct) results. For example, DNS protocol will always give the IP (Internet Protocol) address of the domain name, unless it malfunctions or is attacked. But the security of these components is vulnerable to the attacks and should be improved as described in \cref{PKI,BGP,DNS}.

A blockchain is a public distributed ledger that can record transactions that are connected using a cryptographic hash function \cite{b1}. The basic functionality provided by a blockchain is a secure mechanism for storing and obtaining data, ordered by the timestamp of each record in the data, in a publicly verifiable and immutable manner. For that reason, in most of the system architectures, blockchain provides a storage mechanism for data collection and consensus among participants. In general, blockchain can help in (1) decentralization, (2) provenance and immutability of data, (3) security, and (4) heterogeneity and programmability.

\textbf{Our contributions:} 
In this paper, we first overview the blockchain technology, then provide a discussion of three widely-employed Internet components (PKI, BGP, DNS) and their security vulnerabilities, afterward showing a detailed explanation of the available blockchain-based solutions, and conclude with a summary and open issues.

%The following enlists our contributions:
\begin{comment}

\begin{itemize}
  \item The overview of blockchain and its applications.
  \item Detailed discussion on Internet components and potential security attacks on them.
  \item Discussed the available blockchain-based solutions.
  \item Enlisted open issues for each of the components.
  \item Gave a summary analysis and conclusion of blockchain-based solutions.
\end{itemize}

\end{comment}

\begin{table}[]
\centering
\caption{Legacy Protocol Performance}
\begin{tabular}{|c|c|c|c|c|}
\hline

\textbf{Protocol}                                                     & \textbf{\begin{tabular}[c]{@{}c@{}}Time per\\ query\end{tabular}}           & \textbf{Security}                                               & \textbf{Correctness} & \textbf{Cost} \\ \hline
DNS                                                                   & 0.048s \tablefootnote{https://wp-rocket.me/blog/test-dns-server-response-time-troubleshoot-site-speed}                                                                  & \begin{tabular}[c]{@{}c@{}}Needs \\ improvement\end{tabular} &  High                 & Low           \\ \hline
BGP                                                                   & \begin{tabular}[c]{@{}c@{}}38s \\ for 100\%\\ propagation\end{tabular}  \tablefootnote{http://www.circleid.com/posts/how\_a\_routing\_prefix\_travels\_through\\\_the\_internet}   & \begin{tabular}[c]{@{}c@{}}Needs \\ improvement\end{tabular} &  High                 & Low           \\ \hline
PKI                                                                    & \begin{tabular}[c]{@{}c@{}} Within few \\ milliseconds\end{tabular} \tablefootnote{https://blogs.technet.microsoft.com/option\_explicit/2012/04/19/validating-a-certificate/}  & \begin{tabular}[c]{@{}c@{}}Needs \\ improvement\end{tabular} &  High                 & Low           \\ \hline
\end{tabular}

\label{LPP}
\end{table}

% \begin{figure}[]
% \centerline{\includegraphics[width=50mm,scale=0.8]{internet.png}}
% \caption{Overall Internet Network.}
% \label{fig}
% \end{figure}

\section{Blockchain Technology}
\label{BT}

A blockchain is a decentralized, distributed, and public digital ledger that records data in the form of transactions across multiple devices to enforce immutability, except with a very small probability of the adversary controlling a large fraction of the processing power, stake, etc. A blockchain can be split into \textit{network}, \textit{consensus}, \textit{storage}, \textit{view}, and \textit{side} planes  \cite{b2}, enabling researchers to work on a single idea while improving the overall blockchain infrastructure. 

% Below we list three of the most important planes. \textbf{Network Plane} is responsible for the distribution of the transactions in the form of messages. \textbf{Consensus Plane} keeps total or partial order on all of the globally validated transactions for processing and keeping a consistent state of the system. Proof of Work (PoW) \cite{b1}, Proof of Stake (PoS) \cite{b31}, Proof of Capacity (PoC) \cite{burstcoin}, and Proof of Validation (PoV) \cite{b32} are some consensus protocols that can be used in this plane. \textbf{Storage Plane} stores the validated data (transactions) for future availability and verification.

\textbf{Hash Function:}
Blockchain is built upon collision-resistant deterministic hash functions that map an arbitrary-length input to a fixed-length n-bit output. The hash function should have the property of collision resistance, meaning that an adversary cannot find two different inputs mapping to the same output in polynomial time. This property is important for the integrity and immutability of blockchain.  %The hash should also be random-looking (so it is not predictable) and consistent (same input should produce the same output).

\textbf{Transactions and Blocks:}
% The transaction is a term used because the first blockchain was used by Bitcoin \cite{b1}, where a transaction is the amount of Bitcoin value transferred from one entity to another containing information about the sender and the receiver. Generally, a transaction is data or information of the variant type and can be created by any participant.
The 'transaction' term was first used by Bitcoin \cite{b1}, where a transaction contains the amount of Bitcoin value transferred between entities and information of the sender and the receiver. Generally, a transaction is data or information of the variant type and can be created by any participant.

\begin{figure}[]
\centerline{\includegraphics[width=\columnwidth]{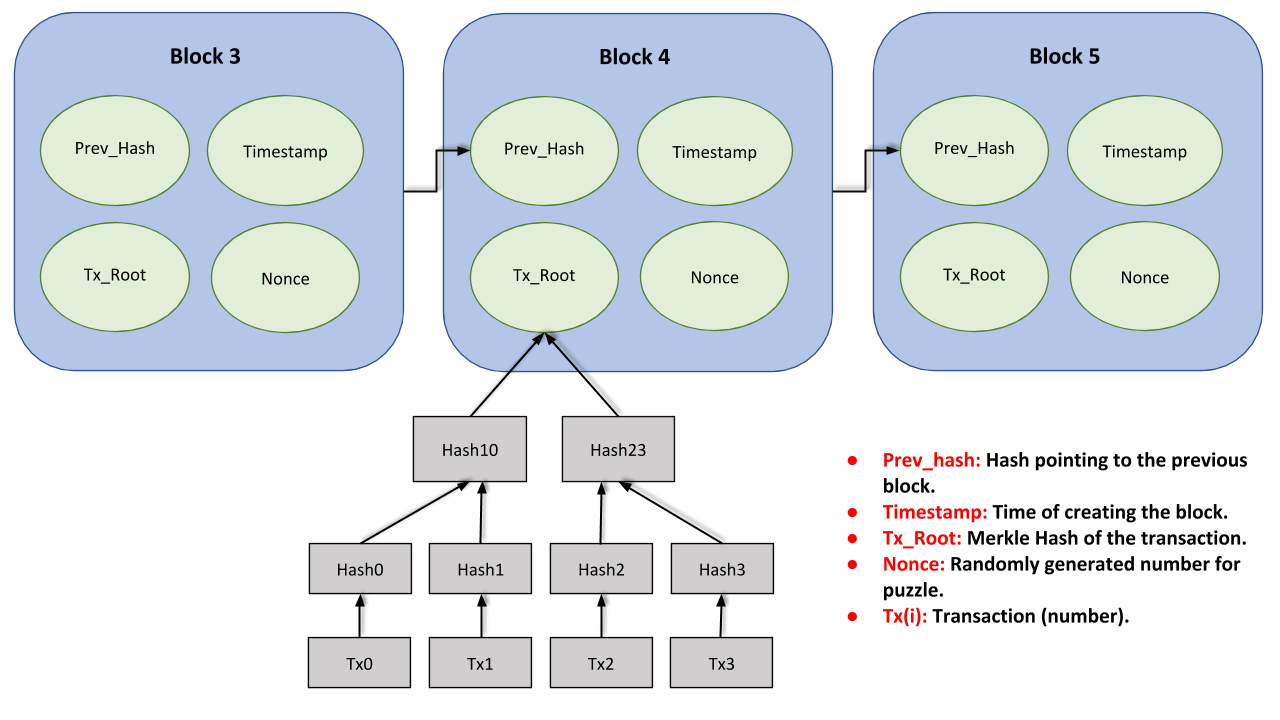}}
\caption{Sample blockchain blocks and transactions.}
\label{fig1}
\end{figure}

Blocks are created by the verifiers (miners). A block is a set of approved transactions, along with a timestamp and a hash pointer to the previous block. The first block of a blockchain is called the Genesis Block. The genesis block is almost always hardcoded with a verifiable universal fact and does not refer to a previous block. As shown in Figure \ref{fig1} hashes of all transactions are kept in a Merkle tree for efficient memory management \cite{b33}. For example \textit{Hash0} is the hash of transaction \textit{Tx0}, \textit{Hash1} is the hash of transaction \textit{Tx1}, \textit{Hash01} is the hash of \textit{Hash0} and \textit{Hash1}, and eventually we have the Merkle root hash \textit{Tx\_Root}. The block also contains the hash of the previous block (\textit{prev\_hash}). Thus, it is computationally infeasible to modify or tamper with the contents of the previous blocks, as this would require finding the hash of all of the remaining blocks to keep the chain connected.

% The block size (the number of transactions in one block) is a critical property of the blockchain as it affects performance and security. Large blocks increase the storage overhead but increase the security, and the inverse can be said for small blocks. Optimizations are done to find a sweet spot between performance and security \cite{b35}. 

\textbf{Types of Blockchains.}
In a blockchain, entities can be readers or writers (writers can be of two types, data owners, who create transactions, and verifiers, who create blocks). Depending on the permissions of these entities, a blockchain can be divided into two groups.
In \textbf{Permissionless Blockchain}, an entity does not require permission to become a reader or writer like Bitcoin \cite{b1} and Zerocash \cite{b36}. Whereas in \textbf{Permissioned Blockchain}, a centralized entity grants permission to the users to be the readers or writers 
% Permissioned blockchains can be divided into public and private permissioned blockchains \cite{b5}. The difference is that in public permissioned blockchains anyone can read but need permission to write. 
(e.g., Hyperledger \cite{b37}).% is an example of permissioned blockchain.
%
% Both types have their advantages and disadvantages depending upon the application. Permissionless blockchains allow more entities to be part of the network but are require a validator to authorize the transaction. This process is time taking and expensive. Also, it is prone to Sybil attacks \cite{b51}, where a network tries to appear as different nodes by creating fake identities. In contrast, permissioned systems use a participant list that speeds up the validation process and decreases the cost of maintenance, but it limits the number of participants in the blockchain. 

\textbf{Consensus Protocols.}
Blockchain presents a solution for the environment where the parties do not have to trust each other and collaborate. As there is no universal trusted third party, each blockchain has to have a consensus protocol for reliability and consistent state of the network.

\textbf{Proof of Work (PoW):} In PoW, a node can get its block accepted if it can solve a cryptographic puzzle (hash) and spend some computational resources in the process. It was first implemented by Bitcoin \cite{b1}.

\textbf{Proof of Stake (PoS):} A node is randomly selected depending upon the stake/resources (ether in Etherium \cite{b31}) she has. Then its block is accepted to be appended to the chain. 
% Transactions as Proof-of-Stake (TaPoS) is a variant where transaction completion is considered as stake \cite{b35}.  

\textbf{Byzantine Fault Tolerance (BFT):} In Practical BFT (PBFT) \cite{pBFT}, there is a leader election (where each entity participates) to elect an entity that has authority to add a new transaction in the chain. This protocol assumes that there more than 2/3 of the honest participants. In Delegated BFT (DBFT) \cite{neo}, participants, by voting, pick the delegate they support. The selected delegates, through the BFT algorithm, reach a consensus and generate new blocks.
% \item \textbf{Other Protocols:} Other used protocols are Proof-of-Activity (PoA) \cite{PoA}, Proof of Burn \cite{Slimcoin}, and Proof of Elapsed Time \cite{SawtoothLake}.

\textbf{Problems.}
Blockchain presents a new perspective on Internet security, but the traditional Bitcoin blockchain (which is used as the solutions discussed in the upcoming sections show) still has issues regarding huge power, energy, storage, and communication requirements \cite{b40} (current Bitcoin blockchain is around 200 GB). Moreover, the Bitcoin blockchain is secure assuming that more than 50\% of the hashing power belongs to the honest parties in the system. There are further attacks on the incentive mechanism, such as selfish mining \cite{b39}.

\section{PKI}
\label{PKI}
Earlier, anyone could pretend to be anybody over the Internet as none of the Internet layers verifies the identity of the entity over the network. This created privacy and trust issues. As a solution, Secure Socket Layer (SSL) and Transport Layer Security (TLS) were introduced. The concept is to provide data integrity, confidentiality, and authenticity using a public/private key pair. But these keys can be compromised. Then another idea was introduced to create cryptographic identities known as digital certificates. Digital certificates contain the identity and public keys of the entity to be used for encryption and authentication. The problem remained that it is very easy to create a digital certificate by self-signing it. There was a need of having a trusted third party that can provide these certificates, now known as a Certificate Authority (CA). CA is a trusted entity that issues digital certificates that verify a digital entity’s identity over the Internet. The infrastructure to manage, store and distribute these certificates is called the Public Key Infrastructure (PKI).

% There is a lot of attacks that are done over CA, SSL and PKI \cite{b18}\cite{b19}, and the researches have been done to mitigate these attacks. In the subsequent subsections, we will discuss in detail these attacks and how Blockchain can help make them more efficient and secure.

%\subsection{Certificate Authority}\label{AA}

%In X.509 public key infrastructure, a certificate authority
A CA signs a certificate to bind the public key of a server to its identity. Then SSL/TLS uses these certificates to authenticate the web-server. Trusting the CA, the browser obtains the server's public key to establish a secure connection. There are two types of certificate authorities \textbf{ROOT-CA} and \textbf{SUB-CA}. The certificate is trusted if it is signed by ROOT-CA. The browsers or operating systems come with many ROOT-CA public keys stored in their databases. As ROOT-CAs might be limited in number and become bottle-neck when there is a lot of demand, there are SUB-CAs that can sign the certificates. PKI works on \textit{Chain of Trust}. To authenticate a certificate, the browser (or any other entity) checks whether or not the certificate is signed by a valid ROOT-CA. If the signer is a SUB-CA, the validation continues in a chain up to the ROOT-CA. Recent research indicates that CAs can be dishonest, get attacked, or can be using faulty or outdated cryptographic algorithms \cite{b3}. Which effects the security of PKI.

% In any of these cases, CA assigns key pairs held by adversarial web-server or entity. Then that adversary can easily harm the security of the network and limits the security of PKI

Currently, there are two types of approaches for the security of PKI.
%\begin{enumerate}
    %\item 
    \textbf{Log Based PKI:} Highly available servers are appointed for publishing and secure monitoring of the certificates to ensure that CAs do not behave maliciously. Still, there are issues with log-based solutions like revocation explained in \cite{b9}.
    %\item 
    \textbf{Web of Trust (WoT):} is a decentralized approach. Users can put their trust in another entity by signing their certificate. Then each trusted entity keeps a certificate that contains signatures of the users that trusted it, in addition to its public key.
%\end{enumerate}

Limitations of PKI security and the advantages of using blockchain to enhance the security and efficacy of PKI is given by \cite{b11}. The architecture assumes the existence of blockchain-backed PKI and uses it to secure the critical (rich) credentials. A privacy-aware PKI system based on blockchain was presented by \cite{b10}. The paper claims that there are many instances (like anonymous social forums), where the entity does not want to reveal its identity. Current PKI leaks this information by knowing which key is used in the protocol. The paper uses blockchain to have online and offline keys and encryption to hide the identity of the user. \cite{b12} proposes Etherium-based blockchain technology to build secure PKI systems, resolving the issues of log-based PKI and the WoT approaches. Blockchain resolves the single point of failure issue and the need for a newcomer to prove its trustworthiness. %The paper gives RESTFul API for registering new CA and adding new certificates while the Validation module is used to validate the chain of trust. The paper analyses the performance and security of the implementation.

\textbf{Certificate Transparency (CT)} was introduced by \cite{b41}, using an append-only public log, to improve the accountability of CAs. As certificates are publicly recorded in the log servers, a fraudulent certificate can be detected, and the countermeasures can be taken to handle the potential attack. Though this removes the central authenticating entity \cite{b41}, considering the increasingly huge number of current certificates, it may introduce computation and communication burden on both clients and servers \cite{b42}. Moreover, it is shown that \textit{split-world} attack can be performed on CT \cite{b43}, where the attacker presents different views of the log to successfully impersonate as the victim. Another issue with CT is that it is not a privacy-preserving scheme \cite{b44}. %The solution for this has been present \cite{b42,b44} which still are prone to the above-mentioned attack.
Certificate Revocation is another phenomenon that reduces the efficacy of these solutions as stale certificates can be used by attackers \cite{b43}. 

\cite{yuce2018server} gives an approach to detect the man-in-the-middle (MITM) attack happened/happening to a victim client. The concept of notary nodes is used, where the server connects and requests the observation of its certificate. In \cite{alnatsheh2018efficient} public notaries are used by the client, hence replacing the dependence on the web browsers/operating systems to validate the certificates. In \cite{yuce2018server}, the MITM can be detected but cannot be prevented, and in \cite{alnatsheh2018efficient}, the users still have to rely on the notaries (which can all be compromised). 

% \begin{figure}[]
% \centerline{\includegraphics[width=70mm,scale=0.8]{pki.jpg}}
% \caption{Public Key infrastructure and SSL/CA.}
% \label{fig}
% \end{figure}

\cite{b3} proposes a blockchain-based solution to construct certificate transparency. The certificates are published as transactions in a global blockchain by the web servers, which is downloaded by browsers. To verify a certificate, the browser just has to see if the certificate is in the blockchain. As the certificates of the authorities are also published, the CAs are also publicly accountable. This creates more trust and decreases the possibility of having a fraudulent CA. Each certificate has a period of validity and can be revoked at will by omitting it from the next block and hence putting into a certificate revocation list (CRL). %The browser can validate the certificate using the public log.
%
%At a high level, web servers publish their certificates as a transaction to the global append-only blockchain using the PoW consensus mechanism. %There are mainly two types of transactions, Type-I is signed by the server using the publishing key. Type-II is used to initialize or reset publishing key pairs. 
%The browser communicates with the peer-to-peer network to download the updated blockchain. % and verifies the certificate before connecting to the server.
%\textbf{Certification and Validation:} This scheme assumes that there are enough honest entities in the network. So the process starts with waiting until enough honest entities have joined the network. Then each one is allocated a publishing key each of which has to be signed by all certifiers in the network. Then this key can be revoked to reset by the G number of certifiers. Using this key, a web-server can append its certificate, which it gets from the CA to the global blockchain using the type-I transaction. The browser validates the certificates by downloading the block headers, which is received after SSL/TLS negotiations and storing it locally. 
%\textbf{Security and Performance Analysis:} 
In this solution, compromised CA can be detected two ways. One is that other CAs will not approve of its transactions. Secondly, the target server will only publish the certificates from valid CAs and the fraudulent certificates will not be appended to the global chain. As the attacker fails to impersonate the target server, a wrong certificate will not be publicized even if the Publishing Key Pair is compromised. In the case both PK and CA are compromised, the countermeasures are taken within a period before the certifiers certify the change. Still, the attacker may prevent the browser from obtaining the blockchain, which is difficult in such a distributed setting. Second, 
there can be forks introduced but unless more than 1/3 of the certifiers are malicious, then this is not possible either \cite{b3}.
However, this scheme has the following drawbacks identified in \cite{b43}:
\begin{enumerate}
    \item An adversary can use unexpired transactions of the revoked certificate to impersonate the victim server; this is a type of a man-in-the-middle attack.
    \item The proposal is inefficient in terms of storage and has large headers.
    \item The proposal depends on the CAs to publish revocation information of the certificate to the blockchain but the compromised CA might not issue Certificate Revocation information to the public.
\end{enumerate}
CertLedger \cite{b43} provides a solution that is resilient to split-world attacks, does not depend on CA for the certificate
revocation, and preserves the privacy of the clients. 

\textbf{Open Issues:} %Paper has following open issues:
% \begin{enumerate}
%    \item  What happens if there are not enough honest entities?
There are still some issues like what is the incentive for the certifiers? What if honest nodes are compromised after joining? What happens when the key pair of the server is compromised? 
% \end{enumerate}

% \begin{figure}[h]
% \centerline{\includegraphics[width=70mm,scale=0.8]{BGP.jpg}}
% \caption{Overview of BGP.}
% \label{fig}
% \end{figure}

\section{Border Gateway Protocol (BGP)}
\label{BGP}
Autonomous Systems (AS) are responsible for the routing among their network typologies. Typically, an AS represents a collection of IP prefixes to which data is routed \cite{b21}. Sometimes these ASes require a flow of data among themselves to reach the destination not present within individual networks. This data flow is peer-to-peer (P2P) in its nature, and is done through the Border Gateway Protocol (BGP) \cite{b49}. In BGP, an AS announces the IP prefix of all the IPs reachable through itself, together with path delay metrics. All the other Internet Service Providers (ISPs) behind ASes update their routing tables according to the BGP announcement. Although efficient
in practice, ASes assume that their neighboring ASes behave honestly and propagate correct routing information (without having the global knowledge). However, the interests of ASes may conflict, and this weak notion of trust can be breached by malicious ASes (e.g. \textit{prefix hijacking} \cite{b21}). 

%Researches work towards providing integrity, confidentiality, authentication, and authorization for routing messages for the security of routing protocols including BGP. There

Previous efforts have developed many solutions like Secure BGP (S-BGP) \cite{b45}, Secure Origin BGP (SoBGP) \cite{b46}, Inter-domain Route Validation (IRV) \cite{b47} and Path-End Validation \cite{b48}. Each of them depends upon a central trusted entry, which has a huge cost of management and is complex. Also, these solutions depend upon the PKI, which has the vulnerabilities described in the previous section. Another problem is the lack of adoption of these protocols as they introduce costly additional infrastructure for their operation. To avoid the cost of adoption, ASes and 
ISPs are reluctant to migrate towards these solutions despite the known 
security threats and their clear benefits.

% \begin{figure}[h]
% \label{BGPA}
% \centerline{\includegraphics[width=80mm,scale=0.9]{BGPAttacks.png}}
% \caption{BGP prefix high-jacking attacks.}
% \label{fig}
% \end{figure}

%\subsection{BGP Attacks}
BGP is open to different kinds of attacks. In \textbf{BGP route manipulation} attack, an adversary manages to change the BGP table to disrupt the traffic of the Internet. Whereas in \textbf{BGP route hijacking}, an attacker AS announces the prefixes belonging to the victim. As a result, the traffic is re-routed to or through the attacker AS. An attacker can also send malicious or faulty BGP traffic to a victim. The victim exhausts its resources to handle the traffic and is left incapable of processing valid BGP traffic. In \textbf{Route Leak}, as the result of a potential attack or the AS malfunction, an AS issues incorrect information about the IP addresses on their network. This results in inefficient routing and failures for the traffic.

\begin{figure}[h]
\centering
\begin{subfigure}{0.40\linewidth}
    \includegraphics[width=\linewidth]{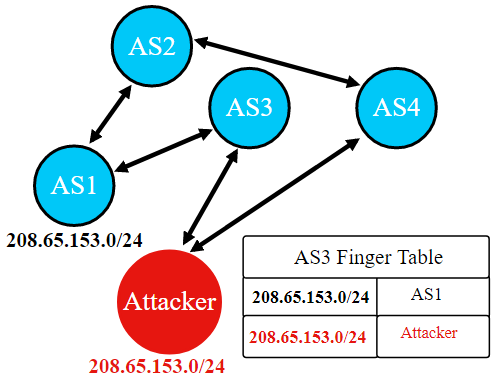}
    \caption{Partial Attack}
    \label{BGPAa}
\end{subfigure}
\hfil
\begin{subfigure}{0.40\linewidth}
    \includegraphics[width=\linewidth]{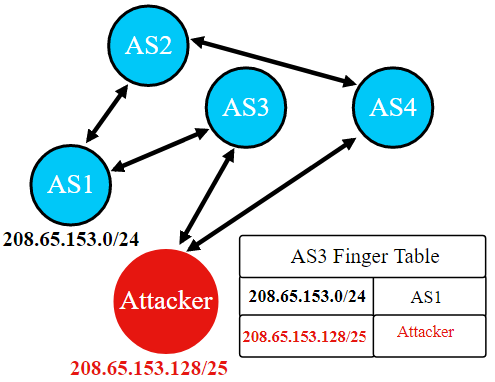}
    \caption{Complete Attack}
    \label{BGPAb}
\end{subfigure}
\caption{Difference between prefix high-jack attacks}
\label{BGPA}
\end{figure}

% Out of them, BGP route hijacking is the most common one \cite{b50}. 
BGP route hijacking is most dangerous and can be classified into two types: partial attack and complete attack \cite{b23}. The partial attack occurs when an adversarial AS announces an identical IP prefix as that of the victim AS. Attack on Youtube in 2008 \cite{b21} is an example of partial hijacking. In a complete attack, the adversary AS announces more specific prefixes than the target AS. Since the default forwarding is based on the longest prefix matching, ASes switch to more specific prefixes and start sending the packet through that route. \cite{b23}. Figure \ref{BGPA} explains the difference between the attacks. In the partial attack, an attacker would announce the 208.65.153.0/24, which is already announced by AS1. Since the two announcements are the same, when any other AS receives the announcement, it can either switch to it or continue with the old routing path. In the complete attack, an attacker would announce 208.65.153.128/25. This IP has a longer prefix match than 208.65.153.0/24 in the respective finger (for example see finger table of AS3 in \ref{BGPAb}) tables so other ASes would switch to this route. 
Interestingly, Youtube used the same concept as a legitimate way to get back the traffic in the 2008 attack \cite{b23}. 
Some BGP attack examples include a global route leak in November 2017, a country-wide Internet outage in Japan due to BGP issues in August 2017, and possible financial traffic re-routing in April 2017.\footnote{https://securityintelligence.com/bgp-internet-routing-what-are-the-threats/}

\begin{comment}

Other examples of BGP attacks are (source: https://securityintelligence.com/bgp-internet-routing-what-are-the-threats/):
\begin{enumerate}
    \item A global BGP route leak, November 2017.
    \item BGP leak incident, Brazil, October 2017.
    \item A country-wide Internet outage due to leaked BGP advertisements, Japan,  August 2017.
    \item A possible rerouted financial network traffic, April 2017.
\end{enumerate}
\end{comment}

\textbf{Blockchain-based Solutions:}
One effort towards securing BGP was to use the PKI presented by IANA \cite{b24} to sign the routes. This scheme is problematic as to effectively use PKI, ASes set up a route assigning authority called Routing Origin Authorization (ROA) which is very costly to implement. Also, as we constructed that PKI infrastructure has security vulnerabilities (section \ref{PKI}), signed BGP update messages based on PKI signatures in BGPSec do not result in a secure path verification protocol.
% because of the security vulnerabilities of the PKI systems discussed earlier

Considering the mentioned problems and the properties of blockchain discussed in Section \ref{BT}, \cite{b24} gives a structure of a system that uses blockchain for the better security and performance of BGP. Blockchain helps in the following ways:
\begin{enumerate}
    \item All transactions occur between peers without any intermediary (like ROA). Also, there is no need for a third party to authentication, eliminating the possibility of tampering or spoofing by a malicious entity.
    \item Provides announcement immutability and re-traceability of the chain of BGP routes. Also, a route is validated by multiple parties and is more trustworthy.
    \item The authors argue that the blockchain should be different from the Bitcoin blockchain and its properties should depend upon the nature of the use case. 
\end{enumerate}
The paper does not provide an implementation and do not discuss the possible attacks and the prevention.
% approach.

The authors in \cite{b21} present a clique-based BGP architecture, RouteChain, to secure BGP against both complete and partial attacks. The method distributes the ASes into subgroups. The system has a global blockchain, and each subgroup has its own private-permissioned blockchain. The main purpose of this sharding is to reduce the storage overhead of having only one global chain and to decrease the transaction validation which is critical for the timely detection of a potential attack. The ASes are grouped based on their geographical proximity for low delays. All groups select a leader to randomly to announce the local routes to the global chain. They use local collaboration among ASes to prevent the complete attack. Whenever there is an update, all the ASes in the group check if their path changes with the update. If it does, they observe the original path and its corresponding prefix. Next, they locate the true owner of the prefix through the global blockchain. If the new update does not belong to the true owner, then the update is discarded. The paper claims that the consensus in the partial attack scenario is achieved in 200 milliseconds, and it is 54.23 seconds for the complete attack. Comparing this timing with the attack on Youtube hijacking incident (in which within 20 minutes, 97 ASes were
hijacked) RouteChain asserts that the system
will notify the ASes about the attack while it is in its initial
stages \cite{b21}. The protocol runs on top of the current BGP architecture, which makes it adaptable and economical.

\begin{comment}
 \cite{b23} and \cite{b26} also give solutions to make RPKI secure. These solutions are valid but the ROA cost described before is still an issue and a hurdle towards the adoption of the system. In contrast, Saad et. al. solution runs on top of the current BGP architecture and is more adaptable.
\end{comment}

\textbf{Open Issues:}
In BGP, different ASes have different policies for sharing the route with the other AS or not. For example, if AS1 does not want its traffic to go through AS2, then AS1 will advertise the path that does not include AS2 even if the resultant path is longer. How can blockchain solutions cater to these 
policies? Can different blockchain architectures be used to solve the BGP security issues?

% \begin{figure}[h]
% \centerline{\includegraphics[width=70mm,scale=0.8]{dnssec.png}}
% \caption{Overview of DNSSec.}
% \label{fig}
% \end{figure}

\begin{table*}%[H]
          \centering
          \captionsetup[subtable]{position = below}
          \captionsetup[table]{position=top}
          \caption{Overall analysis of blockchain based solutions}
          \resizebox{\textwidth}{!}{
          \begin{subtable}{0.5\textwidth}
              \centering
               \begin{adjustbox}{width=\columnwidth,center}
\begin{tabular}{|c|c|c|c|c|c|}
\hline
\textbf{Protocol} & \textbf{Solutions}  & \textbf{Hashed Data}                          & \textbf{Security} & \textbf{Performance}  & \thead{\textbf{Resources} \\ \textbf{Constraints}}                                 \\ \hline
PKI                                                      & \begin{tabular}[c]{@{}c@{}} \cite{b5}, \cite{b15}, \cite{b14},\\ \cite{b16}, \cite{b43} \end{tabular}  & TLS Certificates   & Improved          & Comparable           & \begin{tabular}[c]{@{}c@{}}Storage and \\ computational costs\end{tabular} \\ \hline
DNS                                                          & \begin{tabular}[c]{@{}c@{}} \cite{b18}, \cite{b19},  \cite{b17}, \cite{b13} \end{tabular}  & Domain Names & Improved          & Comparable           & \begin{tabular}[c]{@{}c@{}}Same as of blockchain\end{tabular}              \\ \hline
BGP                                                          & \cite{b22}, \cite{b24}, \cite{b26}   & Domain Routes & Improved          & Comparable           & \begin{tabular}[c]{@{}c@{}} Storage and speed                  \end{tabular}  \\ \hline
\end{tabular}
\end{adjustbox}
                \caption{Blockchain based solutions summary}
                \label{OA}
          \end{subtable}%
          \hspace*{2em}
          \begin{subtable}{0.3\textwidth}
              \centering
              \begin{adjustbox}{width=\columnwidth,center}
                \begin{tabular}{|r|c|c|}
                \hline
                \multicolumn{1}{|c|}{\textbf{Security Features}} & \textbf{Blockchain} & \textbf{Legacy Systems} \\ \hline
                Integrity                                        & High                & Medium                  \\ \hline
                Availability                                     & High                & Medium                  \\ \hline
                Confidentiality                                  & Low                 & Medium                  \\ \hline
                Fault Tolerance                                  & HIgh                & High                    \\ \hline
                No. of Trustless Nodes                           & High                & Low                     \\ \hline
                \end{tabular}
                \end{adjustbox}
                \caption{Security Comparison}
                \label{BVL}
          \end{subtable}
          }
      \end{table*}

\section{Domain Name System (DNS)}
\label{DNS}
% The domain name system was developed during earlier times of the Internet when the use of the Internet was limited. Nowadays, 
The domain name system is used for the resolution of the global domain names. It has a distributed hierarchical design with one root server and 13 specialized servers operated by agencies within a few parts of the world. This design made the system simple, scalable, and flexible. The (in)security of DNS leads to many advanced attacks on DNS including DNS spoofing/cache poisoning, DNS hijacking, and DNS rebinding. A DNS DDoS attack in October 2016 brought down many of the websites including Netflix, Twitter, and CNN \cite{b13}.

%\subsection{DNSSec}

DNSSec is crucial for providing origin authentication and message integrity to communication between the user and the name server. In DNSSec, the root server sends the certificate containing the public key of the next name server along with the IP address of the next server and the hash of the entire message (for data integrity). The next server does the same until the IP address for the domain name is found. This protocol solves the problem of authentication and integrity but still suffers from various attacks namely IP fragmentation and DDoS. The most major problem is its slow adoption.

\textbf{Blockchain-based DNS Alternatives:}
%To deal with the above-mentioned problems, the blockchain alternatives were presented. Some of them, we will discuss here.
Namecoin \cite{b14} was built with the motivation of removing managing domains to avoid trust in a single entity. 
% Namecoin was rather designed as a more general name-value resolving system rather than a substitute for the current DNS system. 
Namecoin uses different prefixes to store and map other types of name-value pairs. For example, the “d/” prefix is used for domain names and “id/” is used to register identities. It further uses the virtual .bit top-level domain name that is not officially registered in the current DNS system. This means Namecoin is isolated from the DNS system and users have to install additional resolving software for resolution of the .bit domain names. Just like DNS, Namecoin provides complete functionalities for registering, renewing, and transferring a domain. The developers modified Bitcoin blockchain to store name-value data like transactions, still utilizing the PoW-based mining mechanism for consensus. Namecoin has shown to have some security flaws: for instance, it was found that a single miner consistently had more than 51\% of the total computing power on the Namecoin network \cite{b15}. In another instance, a Namecoin bug allowed people to steal names from anyone \cite{b15}. Performance-vise, \cite{b15} experienced a latency spike and throughput drop due to software issues of Namecoin. 
% From the business viewpoint, the domain name buyer has no way to regain the investment by returning the name to the open market \cite{studyOfNamecoin}.

Blockstack \cite{b15} combines a DNS system with PKI and purely works with the Bitcoin blockchain. To improve the efficiency of the Bitcoin blockchain for handling a large amount of name-value pairs, a separate logical layer, namely virtual chain, is proposed that works on top of the blockchain to maintain the naming system while the underlying blockchain is only used for achieving consensus on the state of the DNS (or any naming system in general) and the integrity of the name-value data records. Blockstack has significant improvements over Namecoin as it increases the data storage capacity considerably and the virtual chain improves the maintenance of the system. 

In \cite{b13}, authors use blockchain to improve the security and performance of DNSSec. The solution decreases the number of keys needed for DNSSec for easy key management and reduction in DNSSec response size. They use the X509 Cloud blockchain network that is used to store X.509 certificates. This allows us to speed up the verification process. 

DecDNS \cite{b16} gives a blockchain based data storage model for DNS. They also build multiple DNS nodes for further decentralization and addressing single-point-of-failure in DNS resolution. They report 0.006025s response for parallel domain name resolution, which satisfies the DNS performance requirements and shows the potential of using the architecture in the real network for domain name resolutions.

\textbf{Open Issues:}
There should be compatibility between the traditional and the blockchain-based DNS architectures for the systems using the conventional methods for DNS, as this leaves performance and security concerns. For example, a query from the blockchain-backed DNS system can be misinterpreted or discarded by a system using a legacy DNS mechanism. How can blockchain-based and traditional DNS mechanisms inter-operate and integrate? This area of the combination of blockchain-based and traditional decentralization mechanisms for name resolution requires more research.

\section{Comparison}

We finish with an analysis of the mentioned blockchain-based solutions using three main parameters. \textbf{Security:} The level of integrity, confidentiality, and availability introduced by the proposed solutions. \textbf{Performance:} The performance of these solutions compared to the legacy solutions. \textbf{Resource Constraints:} The speed, storage, and cost constraints involved. Table \ref{OA} gives an overview of the proposed solutions per-protocol regarding the analysis parameters.

In Table \ref{BVL}, we identified the key security aspects between blockchain in general and the currently deployed protocols. Blockchain provides Data integrity by the use of hashing in block construction backed by the public key cryptography (as discussed in Section \ref{BT}) to create trust between trustless entities under the same system.. The blockchain excels in availability as the ledger is distributed globally on different nodes offering robustness against the \textit{single point of failure} problem. Further, blockchain provides fault tolerance, i.e., even if some participants leave the system, fail, or get attacked, the blockchain system is not affected as each participant has (more or less) the same copy of the ledger.

% \item \textbf{Energy and Cost:} Bitcoin blockchain requires huge computational power and storage that requires lots of machines and costs more than desirable \cite{b40}. 
%     \item \textbf{Security:} Bitcoin blockchain is secure assuming that more than 50\% of the hashing power belongs to the honest parties in the system. If contrary, the blockchain can be attacked using selfish mining \cite{b39}.
%     \item \textbf{Performance:} Bitcoin blockchain has a massive overhead of hashing and verification. Furthermore, it might take hours or days to get the whole blockchain (currently around 200 GB in size) locally on all the devices. It may even be prohibitively expensive for devices with limited local storage and strict Internet quotas. 

Looking at the drawbacks, we can see that the contents of the records in blockchain are transparent and costs confidentiality to the users considering to opt for blockchain solutions. Furthermore, as discussed in \cite{b40}, blockchain requires lots of storage and computational resources, which come at a high cost and decreased scalability. Furthermore, the blockchain can be attacked (for example using selfish mining \cite{b39}). With the ever-increasing use of the Internet, this issue needs to be solved using better optimizations of currently available blockchain systems, such as LightChain \cite{b32}.

\section{Conclusion}

Blockchain is an emerging solution for improving the security of the overall structure of the Internet to ensure the interruption-free performance of the network. In this paper, we presented blockchain-based solutions for three core network components and systems: PKI, BGP, and DNS, and identified open problems. We plan to further extend our cryptographic discussion with the knowledge we gained from \cite{cryptocourse}.

\begin{comment}

\section{Related Work}

Up to my knowledge, there is no such kind of review that covers all of the protocols of the Internet which use Blockchain. W Meng et. al \cite{b27} gives a review for Intrusion Detection but does not talk about some recent papers. HU Wei-hong et. al. \cite{b28} give a review on blockchain-based DNS.

\end{comment}

\section*{Acknowledgements}
The authors acknowledge TÜBİTAK (the Scientific and Technological Research Council of Turkey) 119E088 grant.

\bibliographystyle{IEEEtran}
\bibliography{IEEEabrv,references}

% \begin{thebibliography}{00}
%     \bibliography{references}
% \end{thebibliography}

\end{document}